\documentclass[letterpaper]{article} 
\usepackage{aaai25}  
\usepackage{times}  
\usepackage{helvet}  
\usepackage{courier}  
\usepackage[hyphens]{url}  
\usepackage{graphicx} 
\usepackage{natbib}  
\usepackage{caption} 
\usepackage{algorithm}
\usepackage{algpseudocode}
\usepackage{amsmath}
\usepackage{amssymb}
\usepackage{bibentry}
\usepackage[inline]{enumitem}
\usepackage{multirow}
\usepackage{booktabs}
\usepackage{array}

\usepackage{newfloat}
\usepackage{listings}
\DeclareCaptionStyle{ruled}{labelfont=normalfont,labelsep=colon,strut=off} 
\floatstyle{ruled}
\newfloat{listing}{tb}{lst}{}
\floatname{listing}{Listing}
\newcommand{\heading}[1]{\vspace*{0.8mm}\noindent\textbf{#1.}}

\title{Attack-in-the-Chain: Bootstrapping Large Language Models for Attacks \\ Against Black-box Neural Ranking Models}
\author {
    Yu-An Liu\textsuperscript{\rm 1,\rm 2},
    Ruqing Zhang\textsuperscript{\rm 1,\rm 2},
    Jiafeng Guo\textsuperscript{\rm 1,\rm 2}\thanks{Jiafeng Guo is the corresponding author.},
    Maarten de Rijke\textsuperscript{\rm 3},
    Yixing Fan\textsuperscript{\rm 1,\rm 2},
    Xueqi Cheng\textsuperscript{\rm 1,\rm 2}
}
\affiliations {
    \textsuperscript{\rm 1}CAS Key Lab of Network Data Science and Technology, Institute of Computing Technology, \\Chinese Academy of Sciences, Beijing, China\\
    \textsuperscript{\rm 2}University of Chinese Academy of Sciences, Beijing, China\\
    \textsuperscript{\rm 3}University of Amsterdam, Amsterdam, The Netherlands\\
    \{liuyuan21b, zhangruqing, guojiafeng, fanyixing, cxq\}@ict.ac.cn, m.derijke@uva.nl 
}

\usepackage{bibentry}

\begin{document}

\maketitle

\begin{abstract}
Neural ranking models (NRMs) have been shown to be highly effective in terms of retrieval performance. Unfortunately, they have also displayed a higher degree of sensitivity to attacks than previous generation models.
To help expose and address this lack of robustness, we introduce a novel ranking attack framework named Attack-in-the-Chain, which tracks interactions between large language models (LLMs) and NRMs based on chain-of-thought (CoT) prompting to generate adversarial examples under black-box settings. 
Our approach starts by identifying anchor documents with higher ranking positions than the target document as nodes in the reasoning chain. 
We then dynamically assign the number of perturbation words to each node and prompt LLMs to execute attacks. 
Finally, we verify the attack performance of all nodes at each reasoning step and proceed to generate the next reasoning step. 
Empirical results on two web search benchmarks show the effectiveness of our method.
\end{abstract}

\section{Introduction}

Neural ranking models (NRMs) are remarkably effective at ranking~\cite{ZhuyunDai2019DeeperTU, guo2016deep, RodrigoNogueira2019PassageRW, yan2021unified,yu2019multi} in information retrieval (IR). But they have also shown vulnerability to carefully crafted adversarial examples~\cite{raval2020one,CongzhengSong2020AdversarialSC,liu2024robust,liu2023robustness}.
This vulnerability raises concerns when deploying NRMs in environments susceptible to black-hat search engine optimization (SEO) \cite{gyongyi2005web}. 
To prevent the exploitation of NRMs, research has focused on the study of adversarial ranking attacks, which aim to deceive NRMs by promoting a low-ranked target document to a higher position in the ranked list for a query through human-imperceptible perturbations \cite{chen2023towards,liu2022order,liu2023topic,wu2022prada}.

Large language models (LLMs) \cite{achiam2023gpt,jiang2023mistral,touvron2023llama,chen2024benchmarking}  have shown strong abilities in understanding, reasoning, and interaction. 
These abilities have enabled LLMs to achieve strong performance in adversarial attacks in natural language processing (NLP) \cite{chao2023jailbreaking,xu2023llm}.
These efforts demonstrate the potential of LLMs to reveal the vulnerability of neural models.
However, adversarial ranking attacks in IR differ from attacks in NLP as they face a multi-step ``battle'' with every top-ranked document in the list to improve their rankings while ensuring efficiency. 
Using the capabilities of LLMs for adversarial ranking attacks remains a challenging and unresolved task. 

Inspired by chain-of-thought (CoT) prompting~\cite{wang2023plan,wei2022chain,xu2023search,cohn2024chain}, 
we develop \textit{attack-in-the-chain} (AttChain), which uses multiple NRM-LLM interaction rounds to effectively achieve attacks. 
Each node in the reasoning chain targets phased ranking improvements, gradually evolving the target document into a fluent and imperceptible adversarial example. 
To this end, we need to resolve two key challenges. 

First, \emph{how to identify nodes in the reasoning chain guiding ranking improvement?} 
We define an \emph{anchor document} to be a document with a higher ranking position than the current perturbed document in the returned list. 
Each anchor document serves as a node to guide the ranking improvement of the target document. 
When the target document achieves a higher ranking, the nodes are updated accordingly to provide further guidance.
Considering all top-ranked documents as nodes increases the computational effort and can be misleading. 
We design a Zipf-based filtering strategy, in which higher-ranked documents are more likely to be retained as candidate documents and selected as anchor documents by LLMs based on previous interactions.
This allows the target document to obtain sufficient informational guidance on its path to improving its ranking.

Second, \emph{how to perturb the target document based on the anchor document node in the reasoning chain?} 
We design a discrepancy-oriented assignment function to dynamically assess the number of perturbation words at each step. 
The key idea is that the degree of perturbation to the target document should be flexibly decided according to its ranking position relative to the anchor document: if the ranking discrepancy is large (small), a large (small) number of perturbation words is needed. 
We then instruct LLMs to generate perturbations via carefully crafted prompts such that the target documents are ranked higher while keeping the perturbations imperceptible. 
Finally, we verify the perturbed documents of all nodes at each reasoning step and select the most effective node to initiate the next attack steps. 
This enables LLMs to dynamically modify the reasoning direction. 

Following \cite{chen2023towards,liu2024multi,wu2022prada}, we focus on a decision-based black-box setting~\cite{brendel2021decision}, where the adversary lacks direct access to model information and can only query the target NRM then receive the ranked list.
We employ GPT-3.5 \cite{chatgpt} and Llama3 \cite{Llama3} as attackers and conduct experiments on two web search benchmark datasets, MS MARCO Document Ranking \cite{nguyen2016ms} and TREC DL19 \cite{Craswell2019TrecDl}. 
The results show that our method significantly outperforms state-of-the-art attack methods in both attack effectiveness and imperceptibility. 
Our proposed method avoids training surrogate models and requires only limited access to the target NRM, thereby reducing the likelihood of detection.
Furthermore, the vulnerabilities of NRMs revealed LLMs can inspire the development of corresponding countermeasures. 

\section{Related Work}

\heading{Adversarial attacks against neural ranking models}
Adversarial ranking attacks are meant to deceive NRMs to promote a low-ranked target document to a higher position in the ranked list produced for a given query by introducing human-imperceptible perturbations \cite{chen2023towards,liu2022order,liu2025robustness,wu2022prada}.
Depending on whether knowledge of the target model can be accessed, the attack task is categorized into white-box and black-box settings \cite{papernot2017practical}.
For practical considerations, existing efforts focus on black-box settings, mainly using the following core steps: 
(i)~training the surrogate model, 
(ii)~identifying vulnerable positions, and 
(iii)~perturbing identified positions~\cite{chen2023towards,wu2022prada,liu2024perturbation}.
Existing work has explored different ways to generate adversarial samples \cite{liu2024robust_survey,liu2025robust}, including the adaptation of textual attacks \cite{liu2022order,wu2022prada}, reinforcement learning \cite{liu2023topic,liu2024multi}, and direct generation using language models \cite{chen2023towards}.
With the emergence of LLMs, the scope of research has expanded to include ranking attacks against LLM-based NRMs \cite{liu2024multi}.
Here, we explore how LLMs can be used to achieve effective attacks against various black-box NRMs. 
The proposed method avoids training surrogate models, reducing access to the target model. 
We hope this method can broaden the understanding and mitigation strategies of such vulnerabilities.

\heading{Adversarial attack with LLMs}
Due to their strong generation abilities, LLMs have been used to conduct adversarial attacks and have demonstrated excellent performance.
\citet{raina2024llm} and \citet{xu2024an} explore using LLMs to generate adversarial examples to attack language models, causing them to produce misleading results.
\citet{gadyatskaya2023chatgpt} and \citet{xu2024autoattacker} investigate using LLMs to plan attack steps and implement automated attacks.
In this work, we study the interaction between LLMs and NRMs, designing chains of reasoning to iteratively perturb target documents with the ultimate goal of maximizing ranking improvements.

\heading{Chain-of-thought prompting}
CoT \cite{wei2022chain} is a technique designed to enhance the reasoning abilities of LLMs by guiding them through a step-by-step process in a few-shot setting.
CoT prompting has proven effective across a variety of tasks, including question answering~\cite{wang2023plan,ji2024chain}, solving mathematical puzzles \cite{giadikiaroglou2024puzzle,madaan2022text}, executing tool calls \cite{paranjape2023art}, understanding graph \cite{liang2023knowledge,liang2024survey}, and evaluating performance \cite{lanham2023measuring}.
In the field of IR, there have been successful attempts by researchers to use CoT to guide retrieval \cite{wang2023self,xu2023search,yu2023improving}. 
Our work diverges from studies that use LLMs to enhance reasoning cooperatively by using CoT reasoning steps for credible, traceable retrieval results. 
Instead, we deploy LLMs adversarially, performing interactive attacks against NRMs in the reasoning tree to explore effective perturbation strategies and generate high-quality adversarial examples.

\begin{figure}[t]
    \centering
    \includegraphics[width=\linewidth]{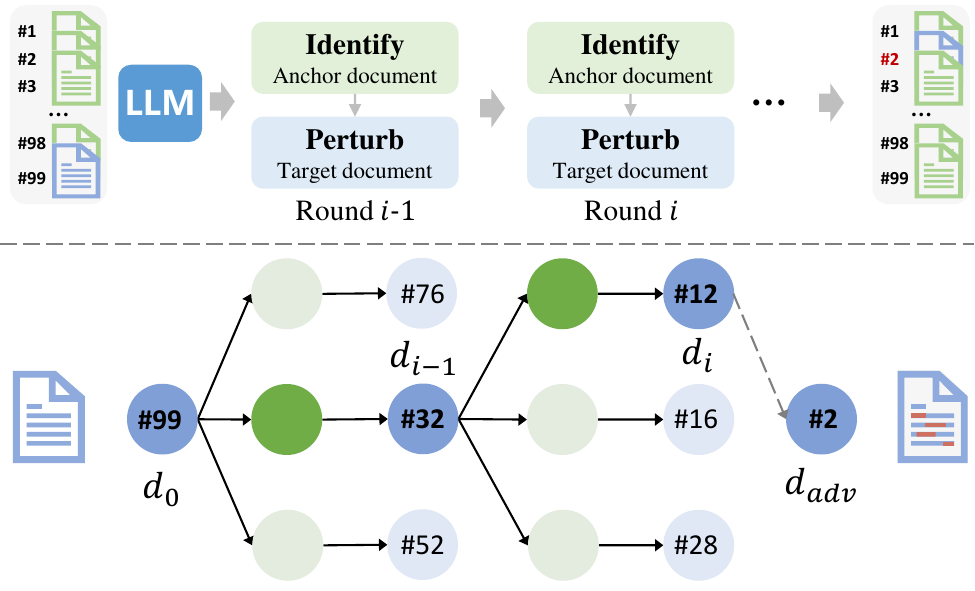}
    \caption{The framework of the proposed method AttChain.}
    \label{fig: attack chain}
    \vspace*{-3mm}
\end{figure}

\section{Method}
As shown in Figure \ref{fig: attack chain}, for each step in the reasoning chain, the proposed attack method AttChain   
\begin{enumerate*}[label=(\roman*)] 
\item first selects several anchor documents as nodes based on the ranking of the current perturbed document (Section~\ref{Sec: selection}); and 
\item then adds perturbations to the target document according to the ranking difference with the anchor document and modifies the reasoning direction (Section~\ref{Sec: stealing}).
\end{enumerate*}
The code is available at \url{https://github.com/Davion-Liu/AttChain}. 

\subsection{Identify nodes} \label{Sec: selection}
\heading{Motivation}
The key idea of this stage is to identify the nodes in the reasoning chain that can guide the LLM to find the direction of boosting the ranking of target document $d$.
A natural idea is to use documents that are ranked higher than the target document, i.e., anchor documents, as nodes to provide attack direction.
Treating \emph{all} top-ranked documents as nodes complicates the reasoning chain, which impairs the clarity of reasoning path and increases computational costs.

A simple method is to directly take the top-1 document as the node.
However, this may not be optimal since it assumes that one document is enough to provide sufficient information to favor an attack.
Ideally, we need a way to filter out a lean and varied candidate set of anchor documents for the LLM to choose from.
Here, we introduce a filter-then-select pipeline to decide the anchor documents on the nodes.

\heading{Zipf distribution-based document filtering}
The aim of this step is to select a set of candidate anchor documents $C_A$, where documents ranked higher are more likely to be retained.
To achieve this, we propose a filtering strategy based on the Zipf distribution~\cite{zipf2016human}, inspired by human click behavior across search engine result pages~\cite{wu2012serial}.
Recall that a Zipf distribution is specified by a rank-frequency distribution where the frequency of any element being sampled is inversely proportional to its ranking. 
The Zipf distribution can be mathematically defined as $P(r; s) \propto r^{-s}$, where $r$ is the rank, and $s$ is the exponent characterizing the distribution.

In the context of our method, at the $i$-th reasoning step, we are given the target NRM $f$, which generates a ranked list $L$ for query $q$, and the perturbed target document at the current step $d_{i-1}$.
$L[:]$ denotes a slice of the list $L$, where $L[:\operatorname{Rank}(f,q,d_{i-1})]$ includes all documents of $L$ up to the rank position of $d_{i-1}$.
Following this, the candidates $C_A$ with $m$ anchor documents are filtered as follows:
\begin{equation}
C_A = \operatorname{Zipf}(L[:\operatorname{Rank}(f,q,d_{i-1})], m, s),
\end{equation}
where $\operatorname{Rank}(\cdot)$ is the ranking position of $d_{i-1}$.

\heading{Anchor document decision prompt}
Given the filtered set $C_A$, our target is to prompt the LLM to select final anchor documents that are considered helpful for the attack. 
Considering input length limitations, for each reasoning step, we 
\begin{enumerate*}[label=(\roman*)]
\item first concatenate the title and the first three sentences of each document in the candidate anchor documents $C_A$; and 
\item then use an \emph{anchor document selection prompt} (Table \ref{fig: prompt}) to organize the text of documents and prompt the LLM to decide $n$ final anchor documents ($n \leq m$) as the nodes.
\end{enumerate*}


\begin{table}[t]
\centering
\scriptsize
\begin{tabular}{|p{7.8cm}|}
\hline
\textbf{Anchor document selection prompt:} \\
You are a search engine optimization specialist aiming to boost the ranking of your target document under the target query. You will receive a target query, a target document, and \(m\) anchor documents. Please select the \(n\) anchor documents that are most useful for improving the target document's ranking under the target query, that is, the ones most worthy of reference. You have moved up \(x\) places in the rankings. Please refer to the previous step and output the id of the anchor documents you have selected and separate the ids by ``\textbackslash n". Follows are target query, target document, and \(m\) anchor documents, give you output: \{Target query\} \{Target document\} \{Anchor documents\} \textbackslash n Output: \\
\vspace{0.5pt}
\textbf{Target document perturbation prompt:} \\
You are tasked as a search engine optimization specialist to enhance the relevance of a target document with respect to a target query. Your goal is to strategically modify the target document to improve its ranking in search results. With the previous step, you have moved up \(x\) places in the rankings. Instructions: \\
1. You are provided with a ``target query", a ``target document", and an ``anchor document". \\
2. Your task is to modify \(|p_i'|\) words in the target document. \\
3. Implement the following strategies: \\
\ \ \ \ a. Identify key phrases or words relevant to the target query from the anchor document. \\
\ \ \ \ b. Combine these key phrases appropriately considering the target query, modify and integrate them into the target document. \\
\ \ \ \ c. Prioritize earlier sections of the document for these changes. \\
4. Please output the perturbed target document in \textless document\textgreater\textless/document\textgreater\ and point out the words you changed and where they are taken from the anchor document: \\
Input: \{Target query\} \{Anchor document\} \{Target document\} \\
Please output the modified target document enclosed in \textless document\textgreater tags: \\
\hline
\end{tabular}
\caption{The anchor document selection prompt and target document perturbation prompts.}
\label{fig: prompt}
\end{table}

\subsection{Perturb documents at nodes} \label{Sec: stealing}
\heading{Motivation}
The key idea is to add perturbations to the target documents based on each node separately at each reasoning step, verifying the most effective node based on the document ranking improvement.
To achieve this, we first assign the number of words to be perturbed in the target document, then we use the LLM to generate the perturbation and update the target document.

\heading{Discrepancy-oriented perturbation assignment}
Intuitively, the gap in ranking position between the target and anchor documents can be bridged by adding perturbations to the target document. 
The larger the gap, the more perturbations are needed. 
Therefore, we design a discrepancy-oriented perturbation assignment method that assigns more perturbations to larger ranking discrepancies while ensuring that the total perturbations remain within the budget. 
Specifically, at the $i$-th reasoning step, the number of words to be perturbed $|p_i^j|$ in the perturbed target document $d_{i}$ according to anchor document $d^j$ is calculated by:
\begin{equation}
|p_i^j| = \frac{\left(\operatorname{Rank}\left(f,q,d_{i}\right) - \operatorname{Rank}\left(f,q,d^j\right)\right)}{\operatorname{Rank}\left(f,q,d\right)}  \ \epsilon ,
\end{equation}
where $\epsilon$ is the budget for the number of manipulated words for the entire attack.

\heading{Target document perturbation prompt}
After obtaining the number of words to be perturbed $|p_i^j|$, we perturb the target document $d_{i-1}$ according to anchor document $d^j$ under the assigned word number $|p_i^j|$. We 
\begin{enumerate*}[label=(\roman*)]
\item first use the \emph{target document perturbation prompt}, shown in Table \ref{fig: prompt}, to guide the LLM in generating perturbations to the target document;  
\item then evaluate the attack effectiveness based on each anchor document at the current nodes; and 
\item finally adopts the node that gives the highest rank improvement and uses it to identify the next round of nodes. 
\end{enumerate*}

The process of identifying nodes and updating nodes is done iteratively. 
During this process, the ranking of the target document is progressively increasing in a ladder-climbing manner. 
Taking into account the computational overhead and effectiveness, the reasoning process executes a total of five rounds.
Then, the final target document is obtained as an adversarial example. 

\section{Experimental Settings}
In this section, we introduce our experimental settings. 

\subsection{Datasets}
\heading{Benchmark datasets}
Following \cite{liu2022order,wu2022prada}, we conduct experiments on two datasets: 
\begin{enumerate*}[label=(\roman*)]
    \item The \textbf{MS MARCO Document Ranking} \cite{nguyen2016ms} (MS MARCO) is a large-scale dataset for web document retrieval, with 3.21 million documents. 
    \item The document ranking task of \textbf{TREC Deep Learning Track 2019} \cite[TREC2019;][]{Craswell2019TrecDl}, which  comprises 200 queries.
\end{enumerate*}

\heading{Target queries and documents}
Following \cite{chen2023towards,liu2022order}, we randomly sample 1,000 Dev queries from MS MARCO and 100 queries from TREC2019 as target queries for each dataset evaluation, respectively.
For each target query, we adopt \textit{Easy} and \textit{Hard} target documents based on the top-100 ranked results from the target NRM.
We randomly choose 5 documents ranked between $[30,60]$ as Easy target documents and select the 5 bottom-ranked documents as Hard target documents.
In addition to the two types, we incorporate \textit{Mixture} target documents for a thorough analysis. These consist of 5 documents randomly sampled from both the Easy and Hard target document sets.

\subsection{Evaluation metrics}
\heading{Attack performance}
We adopt three types of metrics:
\begin{enumerate*}[label=(\roman*)]
\item Attack success rate (ASR) (\%), which evaluates the percentage of target documents successfully boosted under the corresponding target query; 
\item Average boosted ranks (\emph{Boost}), which evaluates the average improved rankings for each target document under the corresponding target query; and
\item Boosted top-10 rate (T10R) (\%), which evaluates the percentage of target documents boosted into the top-10 under the corresponding target query.
\end{enumerate*}
The attack performance of an adversary is better with a higher value for all three metrics. 

\heading{Naturalness performance}
We use five metrics:
\begin{enumerate*}[label=(\roman*)]
\item \emph{Qrs}, which is the average number of queries to the target NRM; 
\item \emph{spamicity detection}, which detects whether target documents are spam;
following \cite{liu2022order,wu2022prada}, we use the utility-based term spamicity detection method OSD \cite{zhou2009osd} to detect the adversarial examples;
\item \emph{grammar checking}, which calculates the average number of errors in the adversarial examples with an online grammatical checker;
following~\cite{chen2023towards,liu2022order}, we use Grammarly\footnote{\url{https://app.grammarly.com/}} for grammar checking; 
\item language model perplexity (\emph{PPL}), which measures the fluency of adversarial examples using the average perplexity calculated using a pre-trained GPT-2 \cite{radford2019language}; and
\item \emph{human evaluation}, which measures the imperceptibility of the adversarial examples following the criteria in \cite{liu2022order,wu2022prada}.
\end{enumerate*}

\subsection{Target NRMs}
Following \cite{chen2023towards,liu2022order,wu2022prada}, we select three typical NRMs as target NRM:
\begin{enumerate*}[label=(\roman*)]
\item BERT;
\item \emph{PROP} \cite{ma2021prop} is a pre-trained model tailored for ranking; and 
\item \emph{RankLLM} \cite{sun-etal-2023-chatgpt} is a model distilled from the ranking capability of an LLM, i.e., ChatGPT into DeBERTa-large \cite{he2020deberta}. 
\end{enumerate*}

\subsection{Baseline methods}
\heading{Baselines}  
\begin{enumerate*}[label=(\roman*)]
    \item \textbf{Term spamming (TS)}~\cite{gyongyi2005web} replaces words with query terms in the target document at a randomly selected position.
    \item \textbf{PRADA}~\cite{wu2022prada} substitutes words in the document with synonym to perform ranking attack against NRMs.   
    \item \textbf{PAT}~\cite{liu2022order} generates a trigger at the beginning of the document for attacks.
    \item \textbf{IDEM}~\cite{chen2023towards} inserts connecting sentences linking original sentences in the document to improve its ranking. 
\end{enumerate*} 

\heading{Model variants} 
We employ the \emph{gpt-3.5-turbo-1106} API provided by OpenAI \cite{openai_api} and \emph{Llama-3-8B} \cite{Llama3} as LLM-based attackers, denoted as \textbf{AttChain$_\mathrm{GPT}$} and \textbf{AttChain$_\mathrm{Llama}$}, respectively. 
Then, based on \textbf{AttChain$_\mathrm{GPT}$}, we consider two variants: 
\begin{enumerate*}[label=(\roman*)]
    \item \textbf{Att}\allowbreak\textbf{Chain$_\mathrm{-CoT}$} rotates the top five documents as anchor documents rather than relying on LLMs to identify them.
    \item \textbf{AttChain$_\mathrm{-dynamic}$} statically adds the same degree of perturbation in each round, regardless of the ranking position gap with the anchor document.
\end{enumerate*}

\subsection{Implementation details}
The initial retrieval step is performed with the BM25 model to obtain the top 100 ranked documents following \cite{liu2023topic,wu2022prada}.
For anchor document filtering, we set $m=20$, $n=5$, and $s=2$. 
For the perturbations, we set the budget for the number of manipulated words $\epsilon$ to 25. 
For human evaluation, we recruit three annotators to annotate 50 randomly sampled adversarial examples and the corresponding documents of each attack method \cite{liu2022order}. 
Following \cite{wu2022prada}, annotators judge whether an example is attacked (labeled as 0) or not (labeled as 1) as the imperceptibility score. 
We repeated our experiment 3 times on 4 $\times$ Tesla V100 32G to get the average results.

\begin{table*}[t]
\centering
\renewcommand{\arraystretch}{1.1}
   \setlength\tabcolsep{6.2pt}
  	\begin{tabular}{l  c c c   c c c    c c c   c c c  }
  \toprule
   & \multicolumn{6}{c}{MS MARCO} & \multicolumn{6}{c}{TREC2019}  \\ 
  \cmidrule(r){2-7} \cmidrule(r){8-13} 
  Method & \multicolumn{3}{c}{Easy} & \multicolumn{3}{c}{Hard} & \multicolumn{3}{c}{Easy}& \multicolumn{3}{c}{Hard} \\
  \cmidrule{1-1} \cmidrule(r){2-4} \cmidrule(r){5-7} \cmidrule(r){8-10} \cmidrule(r){11-13}

   \textbf{BERT}    & ASR & Boost & T10R & ASR & Boost & T10R & ASR & Boost & T10R & ASR & Boost & T10R \\ 
       \midrule
TS & 100.0 & 38.1 & 84.3 & 89.5 & 68.2 & 23.6 
& 100.0 & 36.2 & 81.0 & 90.5 & 65.9 & 21.8 \\
PRADA  & \phantom{1}98.3 & 26.1 & 69.3 & 78.9 & 55.9 & \phantom{1}9.6
& \phantom{1}97.6 & 24.8 & 66.9 & 77.1 & 53.9 & \phantom{1}8.2 \\
PAT   & 100.0 & 35.1 & 78.1 & 82.3 & 60.3 & 18.3 
& 100.0 & 34.3 & 75.6 & 78.3 & 54.1 & 14.9 \\
IDEM   & 100.0 & 39.6 & 85.6 & 90.2 & 69.6 & 25.8
& 100.0 & 37.1 & 82.6 & 87.2 & 65.2 & 22.1 \\
\midrule
AttChain$_\mathrm{Llama}$  & 100.0 & 42.1\rlap{$^{\ast}$} & 92.3\rlap{$^{\ast}$} & 99.1\rlap{$^{\ast}$} & 86.2\rlap{$^{\ast}$} & 34.0\rlap{$^{\ast}$} 
& 100.0 & 40.1\rlap{$^{\ast}$} & 87.2\rlap{$^{\ast}$} & 98.6\rlap{$^{\ast}$} & 84.0\rlap{$^{\ast}$} & 33.1\rlap{$^{\ast}$} \\
AttChain$_\mathrm{GPT}$ & 100.0 & \textbf{44.5}\rlap{$^{\ast}$} & \textbf{94.9}\rlap{$^{\ast}$} & \textbf{99.6}\rlap{$^{\ast}$} & \textbf{91.2}\rlap{$^{\ast}$} & \textbf{39.1}\rlap{$^{\ast}$} 
& 100.0 & \textbf{42.4}\rlap{$^{\ast}$} & \textbf{89.2}\rlap{$^{\ast}$} & \textbf{99.2}\rlap{$^{\ast}$} & \textbf{89.8}\rlap{$^{\ast}$} & \textbf{38.0}\rlap{$^{\ast}$} \\
AttChain$_\mathrm{-CoT}$  & 100.0 & 37.1 & 85.2 & 94.2 & 67.4 & 23.0
& 100.0 & 33.1 & 78.9 & 95.3 & 64.0 & 21.9 \\
AttChain$_\mathrm{-dynamic}$  & 100.0 & 41.8\rlap{$^{\ast}$} & 94.5\rlap{$^{\ast}$} & 99.2\rlap{$^{\ast}$} & 84.3\rlap{$^{\ast}$} & 32.5\rlap{$^{\ast}$} 
& 100.0 & 40.1\rlap{$^{\ast}$} & 88.3\rlap{$^{\ast}$} & 98.2\rlap{$^{\ast}$} & 84.5\rlap{$^{\ast}$} & 33.3\rlap{$^{\ast}$} \\

\midrule
  
   \textbf{PROP}  & ASR & Boost & T10R & ASR & Boost & T10R & ASR & Boost & T10R & ASR & Boost & T10R \\ 
       \midrule
TS & 100.0 & 37.6 & 83.0 & 89.7 & 67.3 & 22.8 
& 100.0 & 34.6 & 79.9 & 91.2 & 65.0 & 20.1 \\
PRADA  & \phantom{1}95.2 & 23.4 & 66.6 & 75.8 & 53.4 & \phantom{1}8.6 
& \phantom{1}94.0 & 21.9 & 62.8 & 72.9 & 52.1 & \phantom{1}6.5 \\
PAT   & \phantom{1}98.6 & 33.6 & 75.9 & 80.2 & 58.7 & 17.3  
& \phantom{1}97.0 & 32.1 & 72.9 & 76.5 & 51.2 & 13.7 \\
IDEM   & 100.0 & 37.3 & 83.0 & 87.9 & 67.5 & 24.0 
& 100.0 & 34.6 & 78.5 & 86.2 & 66.0 & 22.1 \\
\midrule
AttChain$_\mathrm{Llama}$  & 100.0 & 41.0\rlap{$^{\ast}$} & 89.6\rlap{$^{\ast}$} & 98.7\rlap{$^{\ast}$} & 84.5\rlap{$^{\ast}$} & 33.2\rlap{$^{\ast}$} 
& 100.0 & 39.7\rlap{$^{\ast}$} & 86.3\rlap{$^{\ast}$} & 96.3\rlap{$^{\ast}$} & 83.9\rlap{$^{\ast}$} & 32.4\rlap{$^{\ast}$} \\
AttChain$_\mathrm{GPT}$ & 100.0 & \textbf{43.0}\rlap{$^{\ast}$} & \textbf{92.8}\rlap{$^{\ast}$} & \textbf{99.2}\rlap{$^{\ast}$} & \textbf{89.3}\rlap{$^{\ast}$} & \textbf{37.5}\rlap{$^{\ast}$} 
& 100.0 & \textbf{41.2}\rlap{$^{\ast}$} & \textbf{88.2}\rlap{$^{\ast}$} & \textbf{97.2}\rlap{$^{\ast}$} & \textbf{88.3}\rlap{$^{\ast}$} & \textbf{36.2}\rlap{$^{\ast}$} \\
AttChain$_\mathrm{-CoT}$  & 100.0 & 35.8 & 80.1 & 93.8 & 65.2 & 22.4
& 100.0 & 32.3 & 76.2 & 93.5 & 62.5 & 20.0 \\
AttChain$_\mathrm{-dynamic}$  & 100.0 & 40.1\rlap{$^{\ast}$} & 90.5\rlap{$^{\ast}$} & 98.4\rlap{$^{\ast}$} & 82.3\rlap{$^{\ast}$} & 31.2\rlap{$^{\ast}$} 
& 100.0 & 39.7\rlap{$^{\ast}$} & 86.3\rlap{$^{\ast}$} & 96.3\rlap{$^{\ast}$} & 83.9\rlap{$^{\ast}$} & 32.4\rlap{$^{\ast}$} \\
\midrule
   \textbf{RankLLM} & ASR & Boost & T10R & ASR & Boost & T10R & ASR & Boost & T10R & ASR & Boost & T10R \\ 
       \midrule
TS & 100.0 & 34.3 & 79.4 & 89.8 & 63.9 & 19.7 
& \phantom{1}98.9 & 30.6 & 71.0 & 86.8 & 57.6 & 19.0 \\
PRADA  & \phantom{1}92.1 & 21.1 & 60.9 & 68.9 & 50.2 & \phantom{1}6.7 
& \phantom{1}88.7 & 19.8 & 59.9 & 72.3 & 47.8 & \phantom{1}5.8 \\
PAT   & \phantom{1}95.6 & 30.2 & 72.1 & 75.6 & 54.3 & 14.9 
& \phantom{1}94.6 & 28.9 & 67.6 & 74.3 & 48.5 & 10.9 \\
IDEM   & \phantom{1}98.9 & 34.8 & 79.2 & 84.8 & 63.2 & 21.8
& \phantom{1}97.3 & 34.2 & 78.5 & 82.1 & 60.9 & 19.3\\
\midrule
AttChain$_\mathrm{Llama}$  & 100.0 & 38.2\rlap{$^{\ast}$} & 82.9\rlap{$^{\ast}$} & 92.9\rlap{$^{\ast}$} & 81.0\rlap{$^{\ast}$} & 27.9\rlap{$^{\ast}$} 
& 100.0 & 36.5\rlap{$^{\ast}$} & 83.6\rlap{$^{\ast}$} & 93.5\rlap{$^{\ast}$} & 81.0\rlap{$^{\ast}$} & 30.9\rlap{$^{\ast}$}\\
AttChain$_\mathrm{GPT}$  & 100.0 & \textbf{40.5}\rlap{$^{\ast}$} & \textbf{87.5}\rlap{$^{\ast}$} & \textbf{95.6}\rlap{$^{\ast}$} & \textbf{84.3}\rlap{$^{\ast}$} & \textbf{31.6}\rlap{$^{\ast}$} 
& 100.0 & \textbf{38.5}\rlap{$^{\ast}$} & \textbf{85.6}\rlap{$^{\ast}$} & \textbf{95.2}\rlap{$^{\ast}$} & \textbf{83.6}\rlap{$^{\ast}$} & \textbf{32.8}\rlap{$^{\ast}$} \\
AttChain$_\mathrm{-CoT}$  & 100.0 & 32.1 & 79.2 & 81.6 & 72.3 & 19.2
& 100.0 & 32.5 & 76.3 & 86.5 & 63.8 & 22.4 \\
AttChain$_\mathrm{-dynamic}$  & 100.0 & 38.3\rlap{$^{\ast}$} & 85.8\rlap{$^{\ast}$} & 92.8\rlap{$^{\ast}$} & 81.2\rlap{$^{\ast}$} & 28.7\rlap{$^{\ast}$}  
& 100.0 & 36.3\rlap{$^{\ast}$} & 83.1\rlap{$^{\ast}$} & 92.3\rlap{$^{\ast}$} & 79.8\rlap{$^{\ast}$} & 30.2\rlap{$^{\ast}$} \\

\bottomrule
    \end{tabular}
       \caption{Attack performance of AttChain and baselines on MS MARCO and TREC2019; $\ast$ indicates significant improvements over the best baseline ($p \le 0.05$). }
   \label{table:Baseline}
\end{table*}

\section{Experimental Results}
In this section, we report the experimental results to demonstrate the effectiveness of our proposed method.

\subsection{Attack evaluation}
First, we evaluate the \emph{ranking performance} of the target NRM over both datasets. 
For MS MARCO, the ranking performance (MRR@10) of BERT, PROP, and RankLLM is 0.385, 0.389, and 0.399, respectively.
For TREC2019, the ranking performance (nDCG@10) of BERT, PROP, and RankLLM is 0.608, 0.622, and 0.646, respectively.


The \emph{attack performance} comparisons between AttChain and the baselines are shown in Table~\ref{table:Baseline}. 
The performance on the Mixture level target documents is shown in Appendix A.
We have the following observations:
\begin{enumerate*}[label=(\roman*)]
    \item Both NRMs are vulnerable to adversarial attacks, while RankLLM has relatively better adversarial robustness.
    This demonstrates that LLMs mitigate the vulnerability of NRMs, as observed in \cite{liu2024multi}. 
    \item The performance of PRADA is not as good as other baselines.
    This attack method does not exploit the understanding capability of the language model, thus making it difficult to continuously optimize the entire target document.
    \item The attack effectiveness of PAT and IDEM indicates that, language models can interact with NRMs to generate effective adversarial examples.
\end{enumerate*}

When we look at AttChain, we find that:
\begin{enumerate*}[label=(\roman*)]
    \item AttChain$_\mathrm{GPT}$ outperforms all baselines, highlighting that LLMs are inherently good attackers of NRMs.
    They can use their powerful reasoning capabilities can fully capture the preferences of NRMs in interactions, followed by generative capabilities to obtain effective adversarial examples.
    \item The advantage of AttChain$_\mathrm{GPT}$ over AttChain$_\mathrm{Llama}$ indicates that larger-scale LLMs not only have stronger reasoning and generative capabilities, but can also better capture the knowledge of NRMs.
    \item The superiority of AttChain$_\mathrm{GPT}$ over AttChain$_\mathrm{-CoT}$ suggests that, for anchor document selection, the highest ranking is not necessarily the most appropriate.
    LLMs can find the most efficient anchor document at the moment and generate the corresponding perturbation through the reasoning chain.
    \item The advantage of AttChain$_\mathrm{GPT}$ over AttChain$_\mathrm{-dynamic}$ suggests that progressive perturbation from coarse-grained to fine-grained can help target documents improve their ranking efficiently.
\end{enumerate*}

\begin{table}[t]
\centering
   \setlength\tabcolsep{2.5pt}
  	\begin{tabular}{l c  c c   c c}
  \toprule
       Method & Qrs & Grammar & PPL & Impercept. & \textit{kappa} \\
       \midrule
Original & - & \phantom{1}59 & 43.8 & 0.89 & 0.48\\
\midrule
TS  & - & \phantom{1}67 & \phantom{1}63.2 & 0.12 & 0.56\\
PRADA & 218.4 & 108 & 102.8 & 0.48 & 0.51 \\
PAT & \phantom{1}74.8 & \phantom{1}83 & \phantom{1}70.8 & 0.53 & 0.65\\
IDEM & \phantom{1}72.4 & \phantom{1}71 & \phantom{1}48.5 & 0.75 & 0.42 \\
AttChain$_\mathrm{GPT}$ & \phantom{1}25.0 & \phantom{1}61 & \phantom{1}38.3 & 0.85 & 0.57\\
\bottomrule
    \end{tabular}
\caption{Average number of queries, online grammar checker, perplexity, and human evaluation results for attacking RankLLM on MS MARCO.}
   \label{table:natural evaluation}
\end{table}

\begin{table}[t]
\centering
   \setlength\tabcolsep{11pt}
  	\begin{tabular}{l c c c c}
  \toprule
       Threshold & 0.08 & 0.06 & 0.04 & 0.02 \\ 
       \midrule
TS  & 39.2 & 51.9 & 64.0 & 90.1 \\
PRADA  & 12.3 & 23.5 & 39.7 & 61.3 \\
PAT   & \phantom{1}9.1 & 14.9 & 25.4 & 48.4 \\
IDEM   & 16.8 & 28.7 & 46.2 & 71.8 \\
AttChain$_\mathrm{GPT}$ & \phantom{1}\textbf{6.3} & \textbf{11.2} & \textbf{19.4} & \textbf{38.2} \\
\bottomrule
    \end{tabular}
   \caption{The spamming detection rate (\%) via a representative anti-spamming method (OSD) for attacking RankLLM on MS MARCO.}
   \label{table:anti-spamming}
\end{table}

\subsection{Naturalness evaluation}

\heading{Average number of queries, grammar checking, PPL, and human evaluation}
Table \ref{table:natural evaluation} shows the results of the average number of queries to NRM, grammar, PPL, and human evaluation.
We take the Mixture target documents of MS MARCO as examples,  with similar findings on other target documents and datasets.
We observe that: 
\begin{enumerate*}[label=(\roman*)]
    \item The Qrs of AttChain$_\mathrm{GPT}$ is significantly lower than other methods as AttChain$_\mathrm{GPT}$ does not need to train surrogate models by accessing the target NRM multiple times, but instead identifies efficient reasoning chains through LLMs.
    This facilitates its avoidance of suspicion.
    \item The synonym substitution attacks (PRADA) are the trigger injection attacks (PAT) and easily detected. 
    This is because they inevitably introduce grammatical errors or awkward expressions.
    \item AttChain$_\mathrm{GPT}$ outperforms the baselines over all the naturalness metrics, demonstrating the power of LLMs in generating imperceptible adversarial examples.
\end{enumerate*}

\heading{Spamicity detection}
Table \ref{table:anti-spamming} shows the automatic spamicity detection results on Mixture documents with similar findings on other target documents.
We observe that: 
\begin{enumerate*}[label=(\roman*)]
    \item Due to the direct introduction of a large number of query terms, TS can easily be detected as spamming, while other methods are relatively free from this disadvantage.
    \item PAT has the lowest detection rate among the baselines because it actively avoids generating query terms when generating triggers.
    \item AttChain$_\mathrm{GPT}$ outperforms the baselines on spamming detection, demonstrating instructing LLMs can generate natural and hard-to-detect adversarial examples.
\end{enumerate*}

\begin{figure}[t]
    \centering
    \includegraphics[width=\linewidth]{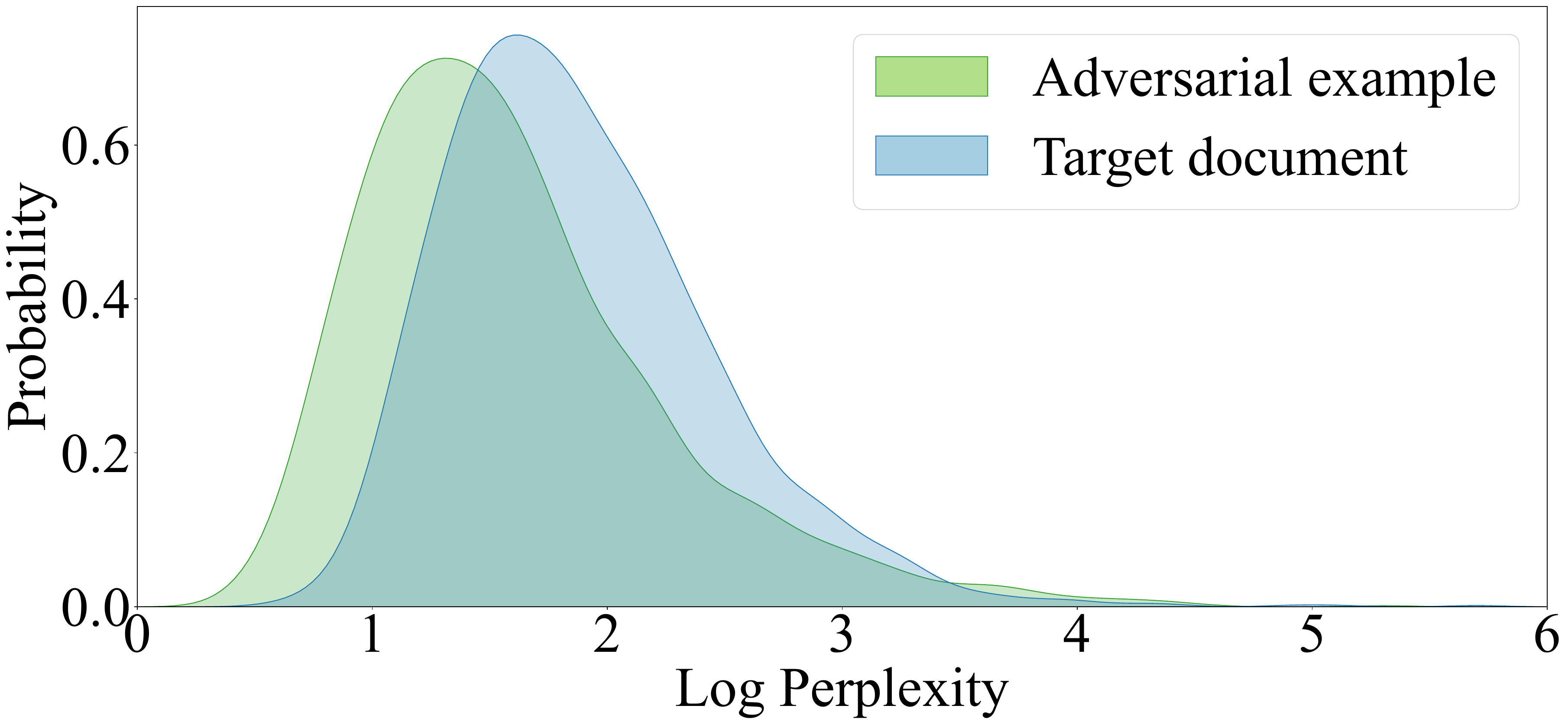}
    \caption{Distributions of log perplexity (PPL) of adversarial examples generated by AttChain$_\mathrm{GPT}$ and target documents on MS MARCO.}
    \label{fig:ppl}
\end{figure}

\begin{figure}[t]
    \centering
    \includegraphics[width=\linewidth]{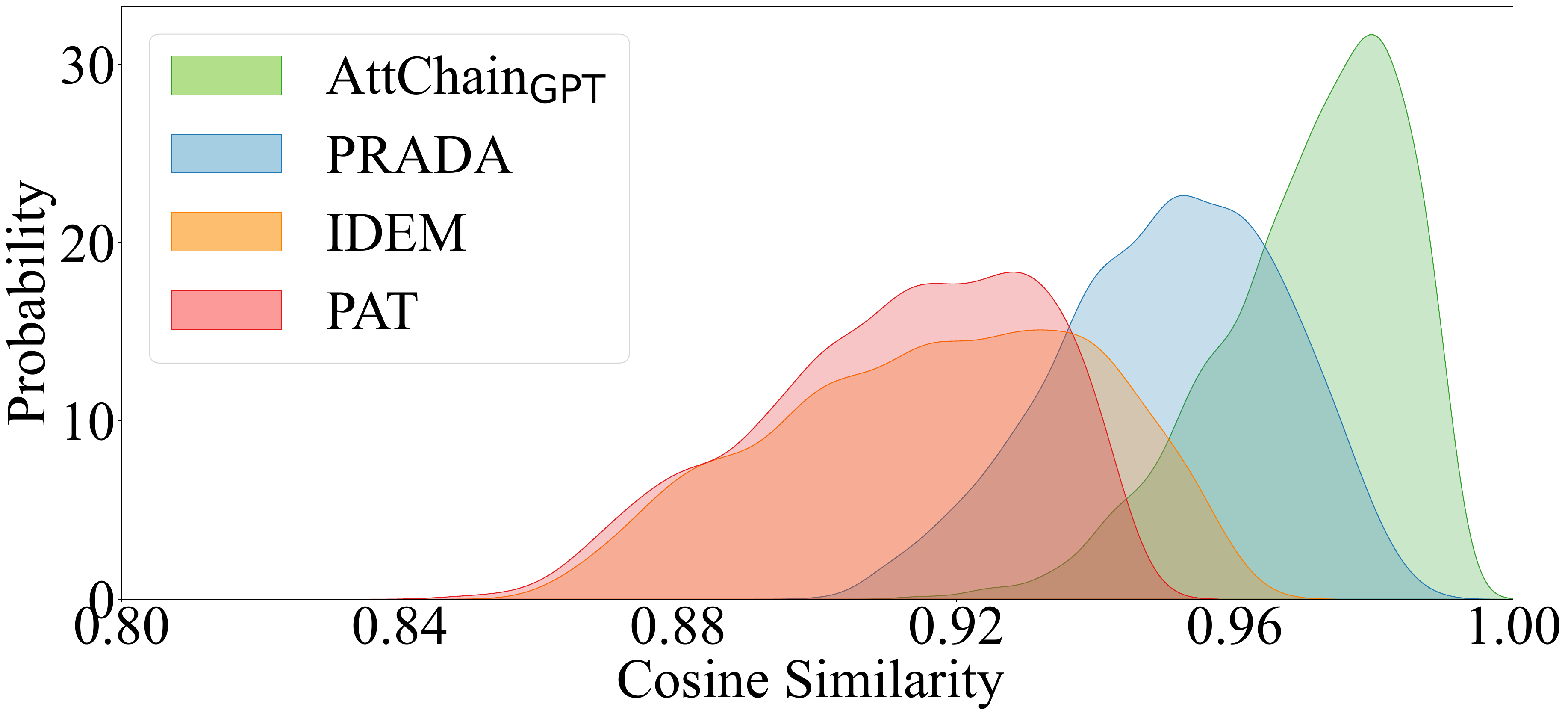}
    \caption{Distribution of cosine similarity of semantic embedding between adversarial examples generated by different attack methods and target documents on MS MARCO.}
    \label{fig:sim}
\end{figure}

\begin{table*}[t]
\centering
\begin{tabular}{l p{13.5cm} c}
\toprule
\textbf{Method} & \textbf{Query:} treating tension headaches medication \quad \textbf{Document title:} Headache Locations Chart & Rank \\
\midrule
\raisebox{-3\height}{Original}  & Headache Locations – What does the location of a headache mean? by Melinda Wilson | Apr 18, 2015 | Headache Locations |Headache is an illness caused by overactivity of, or problems with, structures in the head that are sensitive of pain. Did you know? That there is an organization which advocates the welfare of headache sufferers? National Headache Foundation has categorized headache as a neurobiological disease. With their 45 years of further research and\ldots & \raisebox{-3\height}{98} \\
\midrule
\raisebox{-3\height}{PRADA} & Headache Locations – What does the \textbf{site} of a headache mean? by Melinda Wilson | Apr 18, 2015 | Headache Locations |Headache is an {ailment} caused by overactivity of, or \textbf{issues} with, structures in the head that are sensitive of pain. Did you know? That there is an \textbf{institution} which advocates the welfare of headache sufferers? National Headache Foundation has \textbf{classified} headache as a neurobiological \textbf{illness}. With their 45 \textbf{year} of further research and\ldots & \raisebox{-3\height}{18} \\
\midrule
\raisebox{-3\height}{PAT} & \textbf{where do pains in head live, headache locations mean what nervous} Headache Locations – What does the location of a headache mean? by Melinda Wilson | Apr 18, 2015 | Headache Locations | Headache is an illness caused by overactivity of, or problems with, structures in the head that are sensitive of pain. Did you know? That there is an organization which advocates the welfare of headache sufferers? National Headache Foundation has categorized headache as a \ldots & \raisebox{-3\height}{26} \\
\midrule
\raisebox{-3\height}{IDEM} & Headache Locations – What does the location of a headache mean? by Melinda Wilson | Apr 18, 2015 | Headache Locations | Headache is an illness caused by overactivity of, or problems with, structures in the head that are sensitive to pain. \textbf{This condition can be triggered by tension factors, including headaches with medications.} Did you know? That there is an organization which advocates the welfare of headache sufferers? National Headache Foundation has categorized\ldots & \raisebox{-3\height}{7} \\
\midrule
\raisebox{-3\height}{AttChain$_\mathrm{GPT}$} & Headache Locations – What does the location of a headache mean? by Melinda Wilson | Apr 18, 2015 | Headache Locations |Headache is an illness caused by \textbf{strain or pressure and} overactivity of structures in the head that are sensitive of pain\ldots  the welfare of headache sufferers? National Headache Foundation has categorized headache as a neurobiological disease \textbf{that requires specific treatments}. With their 45 years of further research \textbf{drug-based strategies} and\ldots & \raisebox{-3\height}{2} \\

\bottomrule
\end{tabular}
   \caption{Adversarial examples generated by AttChain$_\mathrm{GPT}$ and other baselines for attacking RankLLM on MS MARCO based on a sampled query with different target documents. Due to space limitations, we only show some key sentences in the document.}
\label{case study}
\end{table*}

\subsection{Mitigation analysis}
We try possible ways of distinguishing the adversarial examples generated by LLMs from the original target document using perplexity (shown in Table~\ref{fig:ppl}) and semantic similarity between origin documents (shown in Table~\ref{fig:sim}).

\heading{Mitigating by perplexity}
We take the adversarial examples and corresponding target documents on the MS MARCO dataset as examples.
Figure \ref{fig:ppl} shows log perplexity distributions, evaluated by GPT-2 \cite{radford2019language} on the two types of documents.
There is a significant distribution overlap between adversarial examples and target documents. 
This implies that filtering adversarial examples through perplexity might result in too many original documents being considered harmful or overlook many adversarial samples that should be detected. 
Therefore, this method seems to struggle to distinguish adversarial samples generated by AttChain from original documents.

\heading{Mitigating by semantic similarity}
we also compare the semantic similarity between the adversarial example and its corresponding original document.
We use the OpenAI semantic embedding \cite{open_embed} to calculate the normalized cosine similarity \cite{rahutomo2012semantic} between the two types of documents.
Figure \ref{fig:sim} shows distributions of cosine similarity across different attack methods. 
We can find that PRADA, due to its use of synonym replacement strategy, generates adversarial samples that have higher semantic similarity to original documents than other baselines.
AttChain generally maintains a high similarity with original documents, because of the effective instruction-following and generation abilities of LLMs. 
This would make it difficult for search engines to distinguish them based on the size of differences during document updates.

On TREC2019 we arrived at similar observations.
Therefore, it is worthwhile to investigate how to recognize adversarial examples through other techniques, such as LLM-generated text detection.
We will further explore the detection of LLM-generated adversarial examples in future work.

\subsection{Case study}
To further understand the proposed AttChain method, we randomly sample a query (ID=524332) and a corresponding hard target document (ID=D1875904) from MS MARCO.
The adversarial examples generated by different attack methods are shown in Table \ref{case study}.
From this example, we find that PAT generates perturbations that are difficult to read, while PRADA may introduce grammatical errors. 
IDEM can generate relatively natural perturbations, but there is a risk of introducing query terms that can easily be detected as spam.
Besides, the adversarial example generated by AttChain$_\mathrm{GPT}$ is more natural-looking that generated by baselines.

\section{Conclusion}
We have proposed an attack method against neural ranking models (NRMs) based on large language models (LLMs).
We employ chain-of-thought (CoT) prompting to perform multiple NRM-LLM interaction rounds in a reasoning chain to generate effective and imperceptible adversarial examples.
Experiments on two web search benchmark datasets show that the proposed method achieves threatening attacks with limited access to NRMs.
Adopting closed-source LLMs, i.e., GPT3.5 improves the attack effectiveness but incurs a relatively large cost.
For future work, we plan to use, and analyze, open-source LLMs to achieve attacks. 

Through our work, we found that LLMs can easily identify NRMs' vulnerabilities in relevance assessment, thereby deceiving NRMs. 
This raises concerns about the use of NRMs in the age of AI-generated content being exploited by search engine optimization (SEO).
We will investigate corresponding defense mechanisms to help develop trustworthy neural IR systems in the future.

\section*{Acknowledgments}
This work was funded by the National Natural Science Foundation of China (NSFC) under Grants No. 62472408 and 62372431, the Strategic Priority Research Program of the CAS under Grants No. XDB0680102 and XDB0680301, the National Key Research and Development Program of China under Grants No. 2023YFA1011602 and 2021QY1701, the Youth Innovation Promotion Association CAS under Grants No. 2021100, the Lenovo-CAS Joint Lab Youth Scientist Project, and the project under Grants No. JCKY2022130C039. 
This work was also (partially) funded by the Dutch Research Council (NWO), under project numbers 024.004.022, NWA.1389.20.\-183, and KICH3.LTP.20.006, and the European Union's Horizon Europe program under grant agreement No 101070212. All content represents the opinion of the authors, which is not necessarily shared or endorsed by their respective employers and/or sponsors.

\bibliography{references}

\end{document}